\documentclass[twocolumn, nofootinbib, nobibnotes, amsmath,amssymb,aps, pra, floatfix]{revtex4-1}

\usepackage{bbm}
\usepackage{mathtools}
\usepackage[export]{adjustbox}
\usepackage{wrapfig}
\usepackage{import}
\usepackage{bbold}
\usepackage{microtype}
\usepackage{bm} 
\usepackage{graphicx}
\usepackage{dcolumn}
\usepackage{color}
\usepackage{silence}

\renewcommand{\v}[1]{\ensuremath{\mathbf{#1}}} 
\newcommand{\gv}[1]{\ensuremath{\mbox{\boldmath$ #1 $}}} 
\newcommand{\uv}[1]{\ensuremath{\mathbf{\hat{#1}}}} 
\newcommand{\abs}[1]{\left| #1 \right|} 
\newcommand{\pd}[2]{\frac{\partial #1}{\partial #2}} 
\newcommand{\grad}[1]{\gv{\nabla} #1} 
\renewcommand{\div}[1]{\gv{\nabla} \cdot #1} 

\begin{document}

\preprint{APS/123-QED}

\title{On the hydrodynamics of Bose-condensed fluids subject to density-dependent gauge potentials.}

\author{Y. Buggy}
 \author{P. {\"O}hberg}
\affiliation{SUPA, Institute of Photonics and Quantum Sciences, Heriot-Watt University, Edinburgh EH14 4AS, United Kingdom}


\begin{abstract}
  When the energy functional of a Bose-condensed state of matter features an effective gauge potential which depends on the density $\rho$ of the condensate, the kinetic energy density of the matter field becomes nonlinear in $\rho$ and additional flow-dependent terms enter the wave equation for the phase of the condensate wavefunction.
  To begin with, we consider a certain class of density-dependent `single-component' gauge potentials, and later extend this class to encompass more general `multi-component' potentials. 
  The nonlinear flow terms are cast into the general form of an inner-product between the velocity field of the fluid and the gauge potential.
  This is achieved by introducing a coupling matrix of dimensionless functions $\gamma_{ij}\left( \rho \right)$, which characterises the particular functional form of the gauge potential and regulates the strengths of the nonlinear terms accordingly. 
  In the momentum-transport equation of the fluid, two non-trivial terms emerge due to the density-dependent vector potential. 
  A body-force of dilation appears as a product of the gauge potential and the dilation rate of the fluid, while the fluid stress tensor features a flow-dependent pressure contribution given by the inner-product of the gauge potential and the current density of the fluid.
  This explicit dependence of the fluid pressure on the flow highlights the lack of Galilean invariance of the nonlinear fluid.

\end{abstract}

\maketitle

\section{\label{sec:intro}Introduction}
In classical mechanics, the interaction of charged particles with the electromagnetic field can be completely described in terms of the force fields $\v{E}$ and $\v{B}$.
The electromagnetic potentials $\phi$ and $\v{A}$ on the other hand, enter merely as auxiliary mathematical quantities bearing no physical significance.
The situation is drastically different in quantum physics: quantisation of a classical theory proceeds from knowledge of the canonical momenta, and it is the energies and momenta which are the central quantities determining the phases of quantum wavefunctions.
As a result, charged particles couple directly to the electromagnetic potentials in the quantum theory, where the form of this coupling notably leads to the Aharanov-Bohm effect and the local gauge invariance of quantum mechanics.
The implications of the fundamental role played by the potentials \cite{aharonov1959significance}, have since led to a diverse range of intriguing physical effects.
These arise through the interplay between particle-particle interactions and applied fields. 
Although the weak field behaviour of gauge-coupled systems is well described by linear response theory, large perturbing field values do not generally allow for a meaningful first order expansion \cite{matt2012}.
As the field is gradually increased, the ordering of the system changes abruptly at certain critical values and a variety of physical phenomena become associated with each intensity range \cite{faze99}: from paramagnetic effects \cite{ston1935}, to the quantum Hall \cite{klitzing1980new,yennie1987integral,pran87} and spin quantum Hall \cite{kane2005quantum,kane2005z,fu2007topological,bernevig2006quantum} effects observed in two-dimensional electron systems.
This notably led to the classification of symmetry protected topological phases of matter \cite{thouless1982quantized,chen2013symmetry} and paved the way for the implementation of topological insulators \cite{qi2011topological,hasan2010colloquium}, illustrating the range of intriguing phenomena which emerge in gauge-coupled many-body systems. 

The charge neutrality of Bose-condensed atomic systems seemingly restricts the discovery of exotic states of matter of this kind.
However, the versatility, controllability and robust character of ultracold quantum gases, have since allowed for the possibility of simulating artificial gauge potentials for charge-neutral systems.
These are generally engineered through combined interactions, such that a system exhibits spatially varying local eigenstates \cite{dalibard2011colloquium,goldman2014light}.
In other words, the action of a gauge potential can be mimicked by imparting a geometric phase onto the wavefunction \cite{berry1984quantal,peskin1989aharonov,dalibard2011colloquium,goldman2014light}.
In this regard, the elucidation of the geometrical nature of the Arahanov-Bohm phase \cite{berry1984quantal} was a landmark in understanding magnetism in quantum mechanics.
Local eigenstates can be induced in a variety of different ways.
Initial attempts exploited the equivalence of the Lorentz and Coriolis forces, by stirring the condensate with a focused laser beam in a magnetic trap \cite{madison2000vortex}, a technique that quickly led to the observation of vortex lattices \cite{abo2001observation}.
More recent implementations have relied almost exclusively on dressing the bare atomic states using light-matter interactions.
For instance, a two-photon Raman scheme \cite{lin2016synthetic} was employed in a series of experiments to engineer both electric \cite{lin2011synthetic} and magnetic \cite{lin2009synthetic} synthetic force fields, as well as synthetic spin-orbit coupling \cite{lin2011spin}, spin Hall effect \cite{beeler2013spin} and partial waves \cite{williams2012synthetic}.
Atomic light-dressing has also opened up the possibility of generating non-Abelian vector potentials with non-commuting components.
These can be implemented for atoms with degenerate eigenstates, and generally emerge when coupling to a laser field produces a degenerate subspace of dressed states \cite{dalibard2011colloquium}.
In addition, efforts have been made to extend the first generation of synthetic potentials - whose space and time dependence are prescribed externally and unaffected by particle motion - and endow these with dynamical properties \cite{lin2016synthetic}. 
For instance, it was recently shown \cite{edmonds2013} how the introduction of weak collisional interactions in an ultracold dilute Bose gas of optically addressed two-level atoms gives rise to a nonlinear effective vector potential $\v{A}(\rho)$ acting on the condensate, where $\abs{\v{A}}$ is modulated by the density of the atomic gas. 
Density-dependent dynamical gauge potentials have also been proposed \cite{greschner2014density} in spin-dependent optical lattices, by combining periodically modulated interactions and Raman-assisted hopping.\\
It should be emphasised that while the nonlinear potential enters the effective gauge-coupled Hamiltonian in the form of a Berry connection, its origin is clearly different from the gauge fields encountered in field theory. 
The degrees of freedom of a gauge field are tied to the points of space and the field evolves under its own equations of motion in accordance with some prescribed Lagrangian density.
In contrast, the ``degrees of freedom'' of a nonlinear synthetic gauge potential are tied to those material elements of a system which are subject to the synthetic potential, by virtue of the combined interactions.
There is no interaction between matter-field and gauge ``field'' since the situation is not that of a dynamical coupling between fields.
Rather, there are not two, but one single field - the matter-field - whose condensed fraction is dynamically governed by a nonlinear wave equation featuring a density-dependent vector potential as a result of the form taken by the Berry connection entering the effective Hamiltonian.\\
It is the emergence of such nonlinear vector potentials which has motivated the present study.
From a hydrodynamical point of view this is an interesting situation because the kinetic energy density becomes nonlinear in the fluid density.
Thus, flow depends explicitly on the density profile of the fluid, where the magnitude of flow of a volume element of fluid typically increases as the volume element shrinks.
A notable consequence of this, is that fluids subject to density-dependent vector potentials lack Galilean invariance, a point which will be covered in greater detail elsewhere.
In this paper, we investigate the types of fluid stress and body-forces which emerge when a quantum fluid couples to nonlinear potentials. 
To undertake this study, we treat the condensate wavefunction as a classical complex scalar field and construct a canonical formalism for the matter-field in terms of the hydrodynamical variables $\rho$ and $\theta$.
Working in this formalism, it will become readily apparent that density-dependent gauge potentials invariably produce flow-dependent terms in the dynamical equation for the scalar velocity potential of the fluid, where such nonlinear terms arise whenever the kinetic energy density depends nonlinearly on the fluid density.
To begin with, we derive a canonical formalism for the Gross-Pitaevskii field and subsequently introduce the nonlinear potentials in the resulting Hamiltonian density through minimal substitution.

\section{\label{sec:canonical_formalism}CANONICAL FORMALISM FOR THE MATTER-FIELD}
\subsection{\label{sec:GP_field}The Gross-Pitaevskii field as a singular Lagrangian system}

At zero temperature, the state of a weakly interacting dilute Bose-Einstein condensed gas is well described by a complex wavefunction $\Psi$. 
The time evolution of the condensate is governed by a nonlinear Schr{\"o}dinger equation (NLSE)
\begin{equation}
i\hbar\partial_t\Psi=\left(-\frac{\hbar^2}{2m}\nabla^2+g\abs{\Psi}^2+V\right)\Psi,
\label{eq:GP_equation}
\end{equation}
known as the Gross-Pitaevskii equation (GPE).
When the number of condensed particles becomes macroscopic, the condensate wavefunction takes on a physical meaning which extends into the classical domain by virtue of the existence of a single phase function common to all particles in the system, a feature which is central to the manifestation of Bose-Einstein condensation and the observation of quantum phenomena at the macroscopic level. 
By treating $\Psi$ as a classical complex scalar field \cite{takabayasi1952formulation,cohen1992atom}, Eq. (\ref{eq:GP_equation}) may be derived by demanding that the action $S\left[\Psi,\Psi^*\right]=\int dt L$, be stationary with respect to variations of the independent fields $\Psi\left(\v{r}\right)$ and $\Psi^*\left(\v{r}\right)$, where the Lagrangian functional
\begin{equation}
  L\left[\Psi,\Psi^*,\dot{\Psi},\dot{\Psi}^*\right]=\int d^3x \mathcal{L}\left(\Psi,\grad{\Psi},\Psi^*,\grad{\Psi}^*,\dot{\Psi},\dot{\Psi}^*\right)
  \label{eq:Lagrangian}
\end{equation}
is given in terms of the Lagrangian density function \cite{holland1995quantum,cohen1992atom,pethick2002bose}
\begin{align}
  \mathcal{L} = \frac{i\hbar}{2}&\left(\Psi^*\dot{\Psi}-\Psi\dot{\Psi}^*\right) \nonumber \\
  &-\frac{\hbar^2}{2m}\grad{\Psi}\cdot\grad{\Psi}^* -\frac{g}{2}\abs{\Psi}^4-V\abs{\Psi}^2.
\label{eq:Lagrangian_GP_Psi}
\end{align}
Denoting the functional derivatives applied to a functional $F\left[\phi\right]=\int d^3x \mathcal{F}\left(\phi,\grad{\phi},\dot{\phi}\right)$ of a field $\phi$, by 
\begin{align}
  \frac{\delta F}{\delta\phi}&=\pd{\mathcal{F}}{\phi}-\grad{\cdot}\pd{\mathcal{F}}{\left(\grad{\phi}\right)}\nonumber\\
  \frac{\delta F}{\delta\dot{\phi}}&=\pd{\mathcal{F}}{\dot{\phi}},
  \label{eq:functional_derivatives}
\end{align}
and inserting Eq. (\ref{eq:Lagrangian_GP_Psi}) into the following Euler-Lagrange field equation:
\begin{equation}
  \frac{\delta L}{\delta\Psi^*}-\pd{}{t}\frac{\delta L}{\delta\dot{\Psi}^*}=0,
\label{eq:EL_GP_Psi*}
\end{equation}
yields the GPE.
Carrying out the same procedure for variation with respect to $\Psi$, yields the complex conjugate of the GPE.

Notice how a complex field variable $\Psi$ automatically requires that $\mathcal{L}$ also depend on $\Psi^*$ in order for the action to be real. 
Yet, the fact that the GPE is first order in time, signifies that $\left(\Psi, \Psi^*, \dot{\Psi},\dot{\Psi}^*\right)$ are not independent and that an excess of dynamical variables are contained in the Lagrangian \cite{cohen1992atom,schiff1968quantum}.
As we shall see in section \ref{sec:constrained_hamiltonian_systems}, this is invariably the situation when the Lagrangian density is linear in the time derivatives of the fields.
However, at first sight this should not come as surprise, since the dynamical state of the field is completely specified by its configuration $\Psi$, whereas the Lagrangian equations of motion are second order in time.
As a consequence, a mechanical field governed by a first order Lagrangian must contain at least two components in order for the configuration of the field to be specified at all times, some combination of these serving as field velocity.
When the dynamical state of a field can be completely specified by a single complex variable $\Psi$, the remaining $6$ real variables contained in $\Psi^*,\dot{\Psi}$ and $\dot{\Psi}^*$, are clearly not dynamically independent of the components of $\Psi$, by definition.
Two equivalent \cite{barcelos1992symplectic} methods have been devised to eliminate redundant variables and construct a reduced phase space for constrained Hamiltonian systems (section \ref{sec:constrained_hamiltonian_systems}): the Dirac-Bergmann method \cite{dirac1958generalized,sundermeyer1982constrained} and the Faddeev-Jackiw method \cite{faddeev1988hamiltonian}.
In the particular case of the Schr{\"o}dinger field, an alternative route is made available by performing a suitable canonical transformation \cite{kerman1976hamiltonian,cohen1992atom}, where one begins by decomposing the field into real and imaginary parts and then supplements the resulting Lagrangian by a total time derivative to obtain a single pair of real conjugate variables.
Several other field transformations yielding Schr{\"o}dinger's equation from a canonical field equation, can also be found in the literature \cite{henley1959elementary,schiff1968quantum}. 
However, these depend either on one or two complex pairs of conjugate variables, meaning redundant variables have not entirely been eliminated.
For a review of these formalisms, and an application of the Dirac-Bergmann and the Faddeev-Jackiw methods to the Schr{\"o}dinger field, see \cite{gergely2002hamiltonian}.
In our study, we take a different approach and derive a canonical formalism for the matter-field in terms of the single pair of real variables $\left(\rho,\theta\right)$, namely the modulus and the argument of the complex field $\Psi$.
These constitute the natural conjugate pair of variables connecting the field and fluid descriptions of a condensate.
It should be emphasised that this approach is by no means original and in fact well-known to classical and quantum hydrodynamics \cite{pokrovskii1976hamiltonian,khalatnikov1978canonical,holm1982poisson,zakh1997hamiltonian,morr1998hamiltonian}, yet perhaps less discussed in the context of field theory.
The line of reasoning followed here may still be appreciated by some readers.

\subsection{\label{sec:constrained_hamiltonian_systems}Constrained Hamiltonian systems}
Nonrelativistic Bose-condensed quantum fluids are generally described by multi-component Lagrangian densities $\mathcal{L}\left(\phi_{\alpha},\grad{\phi_{\alpha}}, \dot{\phi}_{\alpha}\right)$ which are linear in the time derivatives of the fields, where $\phi_{\alpha}\equiv\phi_1,\cdots,\phi_n$.
In this situation, the total Lagrangian of the system can be written as
\begin{equation}
  L\left[ \phi_{\alpha},\dot{\phi}_{\alpha} \right]=\int d^3x\sum_i\mathcal{A}_i\left(\phi_{\alpha},\grad{\phi}_{\alpha}\right)\dot{\phi}_i-V \left[\phi_{\alpha}\right],
  \label{eq:Lagrangian_density_first_order_in_velocities}
\end{equation}
where $V$ is an interaction functional of the field components $\phi_{\alpha}$:
\begin{equation}
  V\left[\phi_{\alpha}\right]=\int d^3x \mathcal{V}\left(\phi_{\alpha},\grad{\phi}_{\alpha}\right).
  \label{eq:interaction_functional}
\end{equation}
The key point to appreciate is that when $L$ is of the form (\ref{eq:Lagrangian_density_first_order_in_velocities}), the canonical momenta are given as functions of $\phi_{\alpha}$ and $\grad{\phi}_{\alpha}$ and as such, can not be treated as independent dynamical variables.  
Indeed, denoting the canonical momentum conjugate to $\phi_i$, by
\begin{equation}
  \pi_i=\pd{\mathcal{L}}{\dot{\phi}_i},
  \label{eq:conjugate_momenta_multi_component_field}
\end{equation}
we notice that if the total Hamiltonian of the system is defined in the usual fashion, according to the Legendre transform
\begin{equation}
  H\left[\phi_{\alpha},\pi_{\alpha}\right]=\int d^3x\sum_{i=1}^n\pi_i\dot{\phi}_i- L\left[\phi_{\alpha},\dot{\phi}_{\alpha}\right],
  \label{eq:Legendre_transform_multi_component_field}
\end{equation}
the field velocities $\dot{\phi}_{\alpha}$ do not appear on the right hand side of this expression as they would typically, given that $\pi_i=\mathcal{A}_i\left( \phi_{\alpha},\grad{\phi}_{\alpha} \right)$ when $\mathcal{L}$ is first order in the $\dot{\phi}_i$.
Thus, the Hamiltonian (\ref{eq:Legendre_transform_multi_component_field}) reduces to the interaction potential functional (\ref{eq:interaction_functional}) and it is not possible to invert $\dot{\phi}_i$ as a function of $\phi_{\alpha}$ and $\pi_{\alpha}$, since 
\begin{equation}
  \frac{\partial^2\mathcal{L}}{\partial\dot{\phi}_j\partial\dot{\phi}_i}=\pd{\mathcal{A}_i}{\dot{\phi}_j}=0,
  \label{eq:non_invertibility_condition}
\end{equation}
holds at every point in space.
Dynamical systems with this property are called ``singular Lagrangian systems'' or ``constrained Hamiltonian systems'' \cite{henneaux1992quantization,sundermeyer1982constrained}.
However, we should keep in mind that the singular nature of a system satisfying (\ref{eq:non_invertibility_condition}), may equally well be the result of actual physical constraints in the system, or through mathematical artefact \cite{faddeev1988hamiltonian}.
Now, although it may not be possible to invert the Legendre transform (\ref{eq:Legendre_transform_multi_component_field}), we can still define the functional
\begin{equation}
  H=\int d^3x\sum_i\pi_i\dot{\phi}_i-L,
  \label{eq:Legendre_transform_1}
\end{equation}
without concerning ourselves for the time being about the details of the Lagrangian density.
In other words, regardless of whether the duality of the Legendre transform holds, we simply observe that Eq. (\ref{eq:Legendre_transform_1}) can be read in reverse, and express the action integral in the canonical form
\begin{equation}
  S=\int dt\left[\int d^3x \sum_i\pi_i\dot{\phi}_i-H\left[\phi_{\alpha},\pi_{\alpha}\right]\right].
  \label{eq:canonical_action_integral_multi_component_field}
\end{equation}
Requiring that Eq. (\ref{eq:canonical_action_integral_multi_component_field}) assume a stationary value for arbitrary variations of the $\phi_i$ and $\pi_i$, yields the system of canonical equations
\begin{align}
  \dot{\phi}_i=&\frac{\delta H}{\delta\pi_i} \label{eq:canonical_multi_component_field_equations_velocities}\\
  \dot{\pi}_i=&-\frac{\delta H}{\delta\phi_i} \quad\quad\quad i=1,\cdots,n \label{eq:canonical_multi_component_field_equations_forces}.
\end{align}
In the case where duality holds, $H\left[\phi_{\alpha},\pi_{\alpha} \right]$ is a functional on the full phase space $\left( \phi_{\alpha},\pi_{\alpha} \right)$ and the $\dot{\phi}_i$ may be inverted to obtain the Lagrangian description of the field.
If, on the other hand, duality does not hold, constraint equations will occur in the system of dynamical equations (\ref{eq:canonical_multi_component_field_equations_velocities}-\ref{eq:canonical_multi_component_field_equations_forces}). 
\subsection{Hydrodynamic canonical formalism}
Following up on discussions in sections \ref{sec:GP_field} and \ref{sec:constrained_hamiltonian_systems}, we choose to describe the matter-field $\Psi$ in terms of the two-component real field $\left(\rho,\theta\right)$, defined according to
\begin{equation}
  \Psi=\sqrt{\rho}e^{i\frac{\theta}{\hbar}},\quad\quad\Psi^*=\sqrt{\rho}e^{-i\frac{\theta}{\hbar}},
\label{eq:polar_form_Psi}
\end{equation}
and treat these components as the independent variables subject to the process of variation \cite{holland1995quantum}.
For a macroscopic occupation of the superfluid state, $\rho=\Psi\Psi^*$ represents the particle density and $\theta$ characterises the superfluid flow.
This `superflow' is of the potential type (\ref{eq:velocity_field_polar_form}), depicting an irrotational velocity field which can be obtained in terms of $\theta$, by substituting Eq. (\ref{eq:polar_form_Psi}) into the current density
\begin{equation}
\v{J}=\frac{i\hbar}{2m}\left(\Psi\grad{\Psi^*}-\Psi^*\grad{\Psi}\right).
\label{eq:current_density}
\end{equation}
This yields the more perspicuous form 
\begin{equation}
\v{J}=\frac{\rho}{m}\grad{\theta}.
\label{eq:current_density_polar_form}
\end{equation}
Given that the fluid occupies a simply-connected region whilst in the superfluid state, Eq. (\ref{eq:current_density_polar_form}) allows for the identification of $\theta/m$ as the potential of the velocity field
\begin{equation}
\v{v}=\frac{\grad{\theta}}{m}.
\label{eq:velocity_field_polar_form}
\end{equation}
Thus, regardless of the physical meaning attributed to the field $\Psi$, there remains from a mathematical perspective, a clear role played by the fields $\rho$ and $\theta$ in establishing matter-wave dynamics: the first of these defines the distribution or amplitude of the matter-field over physical space, while the second dictates the flow of this distribution. 

The canonical field Eqs. (\ref{eq:canonical_multi_component_field_equations_velocities}) and (\ref{eq:canonical_multi_component_field_equations_forces}) for the hydrodynamical variables, are
\begin{align}
  \dot{\rho}&=\frac{\delta H}{\delta\pi_{\rho}} &\dot{\theta}&=\frac{\delta H}{\delta\pi_{\theta}}\nonumber\\
  \dot{\pi}_{\rho}&=-\frac{\delta H}{\delta\rho} &\dot{\pi}_{\theta}&=-\frac{\delta H}{\delta\theta}.
  \label{eq:canonical_equations_hydrodynamic_variables}
\end{align}
Substituting (\ref{eq:polar_form_Psi}) into Eq. (\ref{eq:Lagrangian_GP_Psi}), the Lagrangian density in terms of the new variables, reads
\begin{equation}
  \mathcal{L}=-\rho\left(\dot{\theta}+\frac{(\grad{\theta})^2}{2m}+\frac{g}{2}\rho+V\right) -\frac{\hbar^2}{8m\rho}(\grad{\rho})^2,
\label{eq:Lagrangian_GP_polar}
\end{equation}
and the conjugate momenta (\ref{eq:conjugate_momenta_multi_component_field}) are found to be
\begin{equation}
  \pi_{\rho}=0,\quad\quad\quad\pi_{\theta}=-\rho.
  \label{eq:conjugate_momenta_2}
\end{equation}
These take the form of constraint equations in the system of Eqs. (\ref{eq:canonical_equations_hydrodynamic_variables}), from which we obtain two independent dynamical equations:
\begin{align}
  \dot{\rho}&=\frac{\delta H}{\delta\theta} \label{eq:canonical_equations_hydrodynamic_conjugate_rho}\\
  \dot{\theta}&=-\frac{\delta H}{\delta\rho}. \label{eq:canonical_equations_hydrodynamic_conjugate_theta}
\end{align}
This highlights the fact that the dynamics occur on the reduced phase space $\left( \rho,\theta \right)$ spanned by a single conjugate pair of dynamical variables, where $\theta$ takes on the significance of canonical momentum conjugate to the field $\rho$.
For the Gross-Pitaevskii field in Eq. (\ref{eq:Lagrangian_GP_polar}), the Hamiltonian density governing the time evolution of the canonical pair, can be obtained from Eqs. (\ref{eq:Legendre_transform_1}) and (\ref{eq:Lagrangian_GP_polar}), taking the form 
\begin{equation}
  \mathcal{H}=\rho\left[ \frac{(\grad{\theta})^2}{2m}+\frac{g}{2}\rho+V\right]+\frac{\hbar^2}{8m\rho}\left(\grad{\rho}\right)^2.
  \label{eq:Hamiltonian_density_GP}
\end{equation}
Inserting this expression into the canonical field Eqs. (\ref{eq:canonical_equations_hydrodynamic_conjugate_rho}) and (\ref{eq:canonical_equations_hydrodynamic_conjugate_theta}), yields respectively, a wave equation for $\rho$, given by
\begin{equation}
  \dot{\rho}+\div{\v{J}}=0,
  \label{eq:continuity}
\end{equation}
and a wave equation for $\theta$, in the form
\begin{equation}
  \dot{\theta}+\frac{1}{2}mv^2+g\rho+V+Q=0,
  \label{eq:Q_H_J_GP}
\end{equation}
where $v=\abs{\v{v}}$ and 
\begin{equation}
  Q=-\frac{\hbar^2}{2m}\frac{\nabla^2\sqrt{\rho}}{\sqrt{\rho}}
  \label{eq:quantum_potential}
\end{equation}
is the quantum potential.
Equation (\ref{eq:continuity}) expresses the conservation of fluid mass and Eq. (\ref{eq:Q_H_J_GP}) takes the form of a quantum Hamilton-Jacobi equation (QHJE) and expresses the conservation of mechanical momentum.
The pair of real coupled Eqs. (\ref{eq:continuity}) and (\ref{eq:Q_H_J_GP}) is entirely equivalent to the complex Eq. (\ref{eq:GP_equation}), where relation (\ref{eq:polar_form_Psi}) provides a mapping between both sets of equations known as a Madelung transformation \cite{madelung1926}.

The connection between the Lagrangian and Hamiltonian hydrodynamical descriptions is established by including the constraints in the Legendre transform. 
The constraint equation $\pi_{\rho}=0$ signifies that $\dot{\rho}$ does not enter $\mathcal{L}$, so that the Lagrangian description is obtained simply by transforming variable $\pi_{\theta}$ into $\dot{\theta}$, according to the usual prescription
\begin{equation}
  L=\int d^3x\frac{\delta H}{\delta\pi_{\theta}}\pi_{\theta}-H.
  \label{}
\end{equation}
The second constraint $\pi_{\theta}=-\rho$, then leads to the following relation between $\mathcal{H}$ and $\mathcal{L}$:
\begin{equation}
  \mathcal{L}=-\rho\dot{\theta}-\mathcal{H}.
  \label{eq:Lagrangian_density_Hamiltonian_density}
\end{equation}
Finally, we note that the canonical Hamiltonian density (\ref{eq:Hamiltonian_density_GP}) simultaneously gives the energy density of the matter-field and generates time evolution of an arbitrary functional $F\left[\rho,\theta\right]=\int d^3x \mathcal{F} (\rho,\grad{\rho},\theta,\grad{\theta})$ on the reduced phase space, through the Poisson bracket
\begin{equation}
  \left\{ F,H \right\}=\int d^3x\left[ \frac{\delta F}{\delta\rho} \frac{\delta H}{\delta\theta}- \frac{\delta F}{\delta\theta} \frac{\delta H}{\delta\rho}\right].
  \label{eq:poisson_bracket}
\end{equation}
The matter-field may then be quantised by promoting the fields to operators and Poisson brackets to commutators.
\subsection{\label{sec:nonlinear_A}The nonlinear gauge potential}
The canonical formalism outlined in the previous section in terms of the conjugate pair $\left(\rho,\theta\right)$, allows for the possibility of introducing additional effective interaction terms in a straight forward manner.
For instance, we can investigate the dynamical properties of Bose condensed fluids which are subject to density-dependent effective scalar and gauge potentials $\eta\left( \rho \right)$ and $\v{A}\left( \rho \right)$.
As we shall see, $\v{A}$ gives rise to an expected nonlinear vector potential in the kinetic term of the dynamical equation for the phase, but also interestingly, to additional nonlinear flow-dependent scalar terms.
This is invariably the situation when fluid flow depends explicitly on $\rho$.

An effective gauge-coupled Hamiltonian density can be obtained for the field by demanding that the energy density be invariant under local gauge transformations.
Since a unitary transformation of the condensate wavefunction is equivalent to a local phase transformation and $\theta$ takes on the physical role of a velocity potential, the central quantities concerned by such explicit invariance requirements will be those pertaining to fluid flow.
Accordingly, the current density now takes the explicitly covariant form
\begin{equation}
  \v{J}=\frac{i\hbar}{2m}\left[\Psi\left(\grad{}+\frac{i}{\hbar}\v{A}\right)\Psi^*-\Psi^*\left(\grad{}-\frac{i}{\hbar}\v{A}\right)\Psi\right].
  \label{eq:current_density_gauge_covariant_Psi_form}
\end{equation}
In terms of the conjugate hydrodynamic variables, the current reads
\begin{equation}
\v{J}=\frac{\rho}{m}\left(\grad{\theta}-\v{A}\right),
\label{eq:current_density_gauge_covariant_polar_form}
\end{equation}
and the velocity field, becomes
\begin{equation}
  \v{v}=\frac{1}{m}\left(\grad{\theta}-\v{A}\right).
  \label{eq:velocity_field_gauge_covariant_polar_form}
\end{equation}
Note the distinction which must now be made between the canonical (potential) flow and the mechanical (physical) flow, due to the influence of a geometric vector potential on the motion of the fluid.
In the canonical flow $\v{u}$, we include the total flow which can be accounted for locally by a phase twist in a suitable gauge, whereas the gauge flow denotes any additional flow contribution from $\v{A}$ which can not be accounted for in the phase without destroying the form of the dynamical equation of the fluid.
For example, if in our choice of gauge the wave equation of a fluid reads
\begin{equation}
  \dot{\theta}+\frac{1}{2}mv^2+\phi=0,
  \label{eq:QHJ_generic}
\end{equation}
where $\v{v}$ includes a constant vector potential $\v{A}=\v{c}$, the flow conveyed to the fluid by $\v{A}$ is purely canonical, since performing the transformations
\begin{align}
  \theta&\rightarrow\theta-\int_\mathcal{C} d\v{r}\cdot\v{c}\nonumber\\
  \v{A}&\rightarrow\v{A}-\v{c}, \nonumber
\end{align}
leaves the form of the wave equation unchanged, but the velocity now takes the purely potential form $\v{v}=\grad{\theta}/m$.

\begin{figure}[ht]
  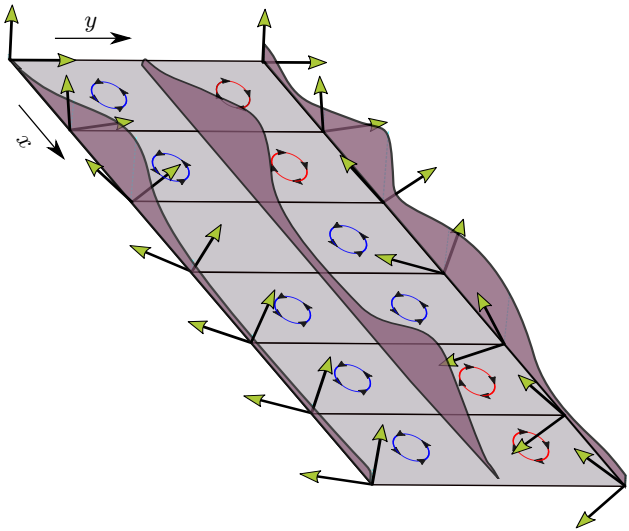
  \caption{A pictorial representation of the twist undergone by the local basis for the phase along two constant-time paths due to a unidirectional density-modulated gauge potential $\v{A}=\rho\abs{\v{a}}\uv{x}$. 
  The waveforms represent three slices of the density $\rho\left( x,y \right)$ in the direction of $\v{A}$.
  The density-dependent gauge potential sets up a gauge flow along $-\v{x}$. 
  Fluid circulation occurs through the spatial dependence of $\rho$ along $y$ (clockwise circulation loops in red and anti-clockwise in blue).
  The basis twist becomes path-dependent in fluid regions with non-vanishing circulation.}   
  \label{fig:basis_twist_gauge_field}
\end{figure}

From Eq. (\ref{eq:velocity_field_gauge_covariant_polar_form}), we see that the transverse component of $\v{A}$ has the effect of biasing or curving solution trajectories, leading to fluid circulation.
Conversely, the longitudinal component tends to stretch or shrink streamlines, leading to fluid dilation.
This is in contrast to the gauge field of standard electrodynamic theory, where only the transverse components enter as dynamical degrees of freedom, the longitudinal component being redundant. 
In the absence of a physical gauge field, the rate at which a fluid flows between neighbouring points of space is dictated by the phase twist: $\v{v}=\v{u}=\grad{\theta}/m$.
The inclusion of a geometric vector potential $\v{A}$ in Eq. (\ref{eq:velocity_field_gauge_covariant_polar_form}) then alters this twist, and does so by rotating or twisting the local basis for the phase along the direction of $\v{A}$ (see FIG.\ref{fig:basis_twist_gauge_field}).
This is an example of a geometric phase being imparted onto the fluid \cite{dalibard2011colloquium,goldman2014light}.
The gauge covariant (mechanical) velocity will be reflected in the kinetic term of the Hamiltonian density, through the mechanical momentum density $\v{g}=m\rho\v{v}$. 
The Hamiltonian density of a quantum fluid subject to nonlinear scalar and gauge potentials, then reads
\begin{equation}
  \mathcal{H}=\rho\left[\frac{1}{2}mv^2+\eta\left( \rho \right)+ V\right] +\mathcal{Q}\left( \rho,\grad{\rho} \right),
  \label{eq:Hamiltonian_density_nonlinear_quantum_fluid}
\end{equation}
where $\v{v}$ from Eq. (\ref{eq:velocity_field_gauge_covariant_polar_form}) features an effective density-dependent vector potential $\v{A}\left( \rho \right)$, $\eta\left( \rho \right)$ is an effective scalar potential, and 
\begin{equation}
  \mathcal{Q}=\frac{\hbar^2}{8m\rho}\left(\grad{\rho}\right)^2,
  \label{eq:quantum_energy_density}
\end{equation}
is the energy density contribution from the quantum potential $Q$ in Eq. (\ref{eq:quantum_potential}).
\section{\label{sec:single_component}The single-component nonlinear gauge potential}
In this paper, the only physical assumption we make is that the basis for $\v{A}$ is externally prescribed and unaffected by the motion of the condensate.
To begin with, we consider the case of a nonlinear gauge potential 
\begin{equation}
  \v{A}=\alpha\left( \rho \right)\v{a},
\label{eq:single_component_nonlinear_gauge_potential}
\end{equation}
whose amplitude is modulated by a density-dependent function $\alpha$ and whose orientation is prescribed by some external vector field $\v{a}(\v{x})$, which we take to be static.
This ensures that the dynamics of the gauge potential depend only on the dynamics of the fluid, but one could easily consider the situation of a time-dependent vector field $\v{a}(\v{x},t)$. 
The configuration of $\v{a}(\v{x})$ would depend on the details of the underlying microscopic model which has produced spatially varying local eigenstates across the system.
In the case of a dilute Bose gas of optically addressed two-level atoms, the introduction of weak contact interactions \cite{edmonds2013} produces spatially dependent perturbed dressed states with an associated density-modulated vector potential $\v{A}=\rho\v{a}$, whose directionality and effective strength are determined by 
\begin{equation}
  \v{a}=\frac{g_{11}-g_{22}}{8\Omega}\grad{\phi},
  \label{eq:static_vector_potential}
\end{equation}
where $\Omega$ is the Rabi frequency characterising the light-matter coupling, $\phi$ is the phase of the laser field and $g_{ij}=4\pi\hbar^2a_{ij}/m$ is the two-body interaction coupling strength associated with the s-wave scattering length $a_{ij}$ between internal atomic components $i$ and $j$.
In this system a geometric phase is acquired by the condensate along constant-time excursion curves through space, whenever the phase of the laser field changes along the curve.
In our study, the nonlinear dependence of the vector potential enters in the form of an arbitrary function of the density $\alpha\left( \rho \right)$.
For convenience, we assume that all the relevent effective strengths have been absorbed into the function $\alpha$, such that $\abs{\v{a}}=1$.
Lastly, although $\v{A}$ is tied to the dynamics of the fluid through $\abs{\v{A}}$, its orientation along $\v{a}$ is fixed in time and as such, possesses only a single dynamical component.
Later, we widen the dynamical scope of $\v{A}$ by considering multiple basis vectors $\v{a}_i$, each with their own $\alpha_i\left( \rho \right)$.
Proceeding in this order, we hope to gain a clear picture of the essential physical features of the elementary case, at which point a mathematical extension of the problem becomes straight forward.
\subsection{Dynamics of the condensate}
The dynamics of the hydrodynamical field variables are governed by the canonical field Eqs. (\ref{eq:canonical_equations_hydrodynamic_conjugate_rho}) and (\ref{eq:canonical_equations_hydrodynamic_conjugate_theta}).
The wave equations for the variables are generated by the total field Hamiltonian obtained from Eq. (\ref{eq:Hamiltonian_density_nonlinear_quantum_fluid}), where $\v{v}$ takes the form
\begin{equation}
\v{v}=\frac{1}{m}\left[ \grad{\theta}-\v{a} \alpha\left( \rho \right)\right].
  \label{eq:velocity_field_single_component_gauge_covariant_polar_form}
\end{equation}
Inserting the resulting Hamiltonian density into Eq. (\ref{eq:canonical_equations_hydrodynamic_conjugate_theta}) again yields an equation of continuity (\ref{eq:continuity}), but with a current density in covariant form.
The second field Eq. (\ref{eq:canonical_equations_hydrodynamic_conjugate_rho}), leads to the QHJE 
\begin{equation}
  \dot{\theta}+\frac{1}{2}mv^2-\gamma\v{v}\cdot\v{A}+\chi\eta+V+Q=0,
  \label{eq:QHJ_single_component_nonlinear_gauge_potential}
\end{equation}
where the density-dependence of the kinetic term within the square brackets of Eq. (\ref{eq:Hamiltonian_density_nonlinear_quantum_fluid}) has produced an additional nonlinear scalar term $-\gamma\v{v}\cdot\v{A}$, where $\gamma$ and $\chi$ are dimensionless functions, defined as
\begin{align}
  \gamma\left( \rho \right)=&\frac{\rho}{\alpha}\frac{d\alpha}{d\rho} \label{eq:gamma_dimensionless_variable} \\
  \chi\left( \rho \right)=&\frac{\rho}{\eta}\frac{d\eta}{d\rho}+1. \label{eq:chi_dimensionless_variable}
\end{align}
These essentially characterise the form taken by the wave Eq. (\ref{eq:QHJ_single_component_nonlinear_gauge_potential}), for given functions $\alpha\left( \rho \right)$ and $\eta\left( \rho \right)$.
For instance, in the simple case where $\alpha\propto\rho^n$ and $\eta\propto\rho^m$, then $\gamma=n$ and $\chi=m+1$ are just numbers regulating the strengths of the flow and density nonlinear terms $\v{v}\cdot\v{A}$ and $\eta$.
More generally however, these will depend on position through $\rho$.
Note that in terms of the logarithmic variable $\tilde{\alpha}=\ln\left({\alpha/\rho}\right)$, $\gamma$ takes the form $\gamma=1+\rho\frac{d\tilde{\alpha}}{d\rho}$.
In the particular case where $\eta=g/2\rho$ and $\alpha=\abs{\v{a}}\rho$ are both proportional to $\rho$ (as seen in \cite{edmonds2013}), the QHJE (\ref{eq:QHJ_single_component_nonlinear_gauge_potential}) together with the equation of mass conservation (\ref{eq:continuity}), map to the NLSE 
\begin{equation}
  i\hbar\partial_t\Psi=\left[ \frac{\left(\v{P}-\v{A}\right)^2}{2m}-\v{a}\cdot\v{J}+g\abs{\Psi}^2+V \right]\Psi,
  \label{eq:GP_equation_single_component_density_modulated_gauge_potential}
\end{equation}
under transformations (\ref{eq:polar_form_Psi}), where $\v{A}=\v{a}\abs{\Psi}^2$ and $\v{J}$ is given by Eq. (\ref{eq:current_density_gauge_covariant_Psi_form}).

In summary, the following key findings may be highlighted.
When a fluid is subject to a nonlinear vector potential, the flow $\v{v}$ in Eq. (\ref{eq:velocity_field_single_component_gauge_covariant_polar_form}) depends explicitly on the density of the fluid such that the kinetic energy density $\kappa=\rho mv^2/2$ becomes nonlinear in $\rho$. 
Thus, the change $\delta\kappa$ in an infinitesimal volume due to $\delta\rho$, is not determined simply by the kinetic energy $mv^2/2$ of the volume as it is typically, but also depends on the overlap between the flow and the vector potential, since $\delta\kappa=\left(mv^2/2+\rho m\v{v}\cdot\partial\v{v}/\partial\rho\right)\delta\rho$.
In the same way that a \textit{scalar potential energy} density with nonlinear $\rho$-dependence $\rho\eta\left( \rho \right)$ produces a \textit{density}-dependent scalar potential $\chi\eta$ in the wave equation under Eq. (\ref{eq:canonical_equations_hydrodynamic_conjugate_theta}), a \textit{kinetic energy} density with nonlinear $\rho$-dependence gives rise to a \textit{flow}-dependent scalar potential. 
This is a general feature which is entirely independent of the details of the microscopic model.
We may also note that the procedure followed in this section outlines a simple method for investigating the implications of introducing effective nonlinear $\rho$-dependent or $\theta$-dependent interaction terms.
One could envisage other forms of coupling between the fields $\rho$ and $\theta$, or introduce additional atomic species, and check whether these give rise to interesting dynamical terms in the resulting wave equation.

\subsection{\label{sec:cauchy}Cauchy's equation}
The remainder of section \ref{sec:single_component} will be devoted to deriving a hydrodynamic Cauchy equation describing the transport of mechanical momentum in the fluid. 
To this end, we begin by stating the general form of this equation and highlight the central quantities relevant to the problem. 
Denoting the time derivative operator in the reference frame of the fluid or \textit{convective derivative} operator, by a total time derivative
\begin{equation}
  \frac{d}{dt}\equiv\pd{}{t}+\v{v}\cdot\grad{},
  \label{eq:convective_derivative}
\end{equation}
and adopting the usual summation convention over indices appearing twice, Cauchy's equation \cite{aris2012vectors} then takes the form 
\begin{equation}
  \rho m\frac{dv_k}{dt}=\rho f_k + \nabla_j \Pi_{jk},
  \label{eq:Cauchy_equation}
\end{equation}
where $f_k$ denote the \textit{body forces} and $\Pi_{jk}$ are the components of the \textit{stress tensor} of the fluid.
Equation (\ref{eq:Cauchy_equation}) holds true for any fluid medium, irrespective of the manner in which stress is connected to the rate of strain.
The distinction between body and stress forces is established according to the manner in which each type act on an infinitesimal volume element of fluid.
The former act throughout the volume thereby changing the overall rate of flow of the volume element, whereas the latter deform the volume element by acting on its bounding surface and lead to propagation of fluid disturbances. 
In other words, the $\Pi_{jk}$ define a linear map between the surface normal vectors and the forces acting on these.
Thus, the stress tensor of a fluid may be written in the form 
\begin{equation}
  \Pi_{jk}=-P\delta_{jk}+\sigma_{jk},
  \label{eq:stress_tensor_pressure_viscocity}
\end{equation}
where $P$ is the fluid pressure associated with normal forces and the $\sigma_{jk}$ account for shearing forces.
It should be emphasised that in accordance with Eqs. (\ref{eq:Cauchy_equation}) and (\ref{eq:convective_derivative}), $P$ does not represent the pressure at a fixed point of space, but the pressure of an infinitesimal volume element which flows with the fluid. 
As a final point, notice how body-forces will typically be the result of the fluid interacting with external fields.
In contrast, fluid stress emerges due to internal-type interactions between the fluid particles, these generally giving rise to nonlinear effective potentials in the wave equation for the fluid. 
However, if the nonlinear potential is not simply a scalar but a tensor of higher rank whose orientation or basis is prescribed externally (\v{a} in this instance), this separation no longer applies.
Thus, $\v{A}\left( \rho \right)$ will be seen to play a double role in Eq. (\ref{eq:Cauchy_equation}), bearing implications for both $\Pi_{jk}$ and $f_k$.

\subsection{\label{sec:momentum_transport_single_component}Momentum-transport from the fluid description of the condensate}
Inspecting the form of Eq. (\ref{eq:velocity_field_gauge_covariant_polar_form}), reveals that the dynamics of the velocity field can be obtained by taking the gradient of the QHJE (\ref{eq:QHJ_single_component_nonlinear_gauge_potential}). 
Doing so, we find that
\begin{widetext}
\begin{equation}
  m\left( \pd{}{t}+\v{v}\cdot\grad{} \right)\v{v}+\pd{\v{A}}{t}+m\v{v}\times\gv{\omega}=-\grad{\left(V+Q+\chi\eta-\gamma\v{A}\cdot\v{v}\right)},
  \label{eq:Cauchy_equation_single_component_nonlinear_gauge_potential_long}
\end{equation}
\end{widetext}
where the kinetic energy term in Eq. (\ref{eq:QHJ_single_component_nonlinear_gauge_potential}) combines to give rise to a convective derivative operator (\ref{eq:convective_derivative}) and also leads to a vortical force due to the non-vanishing fluid vorticity 
\begin{equation}
  \omega_k=\epsilon_{ijk}\nabla_iv_j,
  \label{eq:vorticity}
\end{equation}
stemming from the rotational component of the vector potential.
Alternatively, if we define a synthetic magnetic field 
\begin{equation}
  B_k=\epsilon_{ijk}\nabla_iA_j,
  \label{eq:synthetic_magnetic_field}
\end{equation}
the vortical force can be given a familiar magnetic form through a trivial substitution of the vorticity for $B_k=-m\omega_k$.
Similarly, the time dependence of the vector potential together with the spatial dependence of the scalar field
\begin{equation}
  \phi=\chi\eta-\gamma\v{v}\cdot\v{A},
  \label{eq:scalar_potential_due_to_nonlinear_potentials}
\end{equation}
define a synthetic electric field
\begin{equation}
  E_k=-\nabla_k \phi-\partial_tA_k.
  \label{eq:synthetic_electric_field}
\end{equation}
Accordingly, the forces appearing in Eq. (\ref{eq:Cauchy_equation_single_component_nonlinear_gauge_potential_long}) as a result of the nonlinear potentials, now take the Lorentz form
\begin{equation}
  \lambda_k=E_k+\epsilon_{ijk}v_iB_j.
  \label{eq:synthetic_lorentz_force_1}
\end{equation}
In turn, this leads to the following equation for the force acting on a infinitesimal volume element which flows with the fluid:
\begin{equation}
  m\rho\frac{dv_k}{dt}=\rho\lambda_k-\rho\nabla_k\left( V+Q \right).
  \label{eq:Cauchy_equation_2}
\end{equation}
It should be emphasised that $E_k$ and $B_k$ have been defined only to make contact with the Lorentz force of electromagnetism, the form of which generally emerges whenever a system is subject to scalar and vector potentials.
On this note, although $\lambda_k$ may appear to take the form of a body-force as seen in Eq. (\ref{eq:Cauchy_equation_2}), the nonlinear character of the potentials leading to Eq. (\ref{eq:synthetic_lorentz_force_1}) suggests that this is not the case.
In fact, $\lambda_k$ will soon be given a different form when it becomes apparent that the density dependence of $\v{A}$ signifies that both $B_k$ and $E_k$ are connected to the flow profile of the fluid, at which point a separation of $\lambda_k$ into body and stress terms will be made.
This connection is already clearly apparent for magnetic forces, since $\v{B}$ is proportional to the vorticity by definition.
For a superfluid confined to a two-dimensional surface which is left to evolve freely from some initial configuration, this implies that the total synthetic magnetic flux is proportional to the fluid circulation on the boundary and therefore conserved during the motion of the fluid.
Thus, a superfluid subject to $\v{A}\left( \rho \right)$ is constrained to evolve within a subspace of density configurations having identical circulation on the boundary.
In the case of a uniform field $\v{a}$, infinitesimal circulation loops arise when $\rho$ is asymmetrically distributed about $\v{a}$ (see FIG.\ref{fig:basis_twist_gauge_field}).
In the more general case of non-vanishing 
\begin{equation}
  b_k=\epsilon_{ijk}\nabla_i a_j,
  \label{eq:magnetic_body_field_single_component}
\end{equation}
$B_k$ is comprised of two parts: 
\begin{equation}
  B_k=\alpha b_k+\frac{\gamma}{\rho}\epsilon_{ijk}A_j\nabla_i\rho,
  \label{eq:synthetic_magnetic_field_two_parts}
\end{equation}
namely, an $\alpha$-modulated body-type magnetic field stemming from the spatial dependence of $\v{a}$, and a stress-like magnetic field associated with density variations transverse to $\v{a}$.

While a transverse flow component can always be attributed to a synthetic magnetic field, the density-dependence of $\v{A}$ connects the synthetic electric field to the longitudinal component of flow.
Indeed, an equation of conservation (e.g. Eqs. (\ref{eq:continuity}, \ref{eq:transport_energy}, \ref{eq:transport_momentum})) relates the time dependence of a physical quantity to the longitudinal component of its current.
For $\v{A}$ in Eq. (\ref{eq:single_component_nonlinear_gauge_potential}), this signifies that
\begin{equation}
  \pd{A_k}{t}=-\frac{\gamma A_k}{\rho}\nabla_iJ_i.
  \label{eq:continuity_gauge_potential}
\end{equation}
This simple yet significant equation, encapsulates the underlying connection between the dynamics of the gauge potential and the dynamical state of the condensate.
As a result, the synthetic electric field defined in Eq. (\ref{eq:synthetic_electric_field}) can be spatially resolved, as
\begin{equation}
  E_k=-\nabla_k\phi+\frac{\gamma A_k}{\rho}\nabla_iJ_i,
  \label{eq:synthetic_electric_field_2}
\end{equation}
indicating that a non-trivial type of force should be expected for the onset of current in the system.
 
Equations (\ref{eq:vorticity}), (\ref{eq:synthetic_magnetic_field}) and (\ref{eq:synthetic_electric_field_2}) suggest that we dispose of the synthetic fields and express the Lorentz force in terms of the symmetric and antisymmetric parts of the velocity gradient tensor or \textit{displacement field} tensor
\begin{equation}
  \nabla_i v_j\equiv d_{ij}=e_{ij}+\Omega_{ij},
  \label{eq:velocity_gradient_tensor}
\end{equation}
where $e_{ij}$ is the \textit{deformation} or \textit{rate of strain} tensor, given by
\begin{equation}
  e_{ij}=\frac{1}{2}\left( \nabla_i v_j + \nabla_j v_i \right),
  \label{eq:strain_tensor}
\end{equation}
and $\Omega_{ij}$ is the \textit{vorticity} tensor
\begin{equation}
    \Omega_{ij}= \frac{1}{2}\left( \nabla_i v_j - \nabla_j v_i \right).
  \label{eq:vorticity_tensor}
\end{equation}
The diagonal components of $e_{ij}$ dictate the rates of longitudinal strain connected with pure stretching whereas the off-diagonal components determine the rates of shear strain connected with pure shearing \cite{aris2012vectors}.
In terms of these objects, the longitudinal component of the velocity field or \textit{dilation rate}, takes the form
\begin{equation}
  \nabla_i v_i=\delta_{ij}d_{ij}=\delta_{ij}e_{ij},
  \label{eq:divergence_velocity_field}
\end{equation}
while the transverse component or vorticity from Eq. (\ref{eq:vorticity}), reads
\begin{equation}
  \omega_k=\epsilon_{ijk}d_{ij}=\epsilon_{ijk}\Omega_{ij}.
  \label{eq:curl_velocity_field}
\end{equation}
After inserting Eq. (\ref{eq:synthetic_electric_field_2}) into Eq. (\ref{eq:synthetic_lorentz_force_1}) and noting that $B_k=-m\epsilon_{ijk}\Omega_{ij}$, we find that the Lorentz force is related to $e_{ij}$ and $\Omega_{ij}$, in the form 
\begin{equation}
  \lambda_k=-\nabla_k\phi+\gamma A_k\delta_{ij}e_{ij}+mv_i\left(2\Omega_{ik}+\frac{\gamma A_k}{m\rho}\nabla_i\rho\right).
  \label{eq:synthetic_lorentz_force_2}
\end{equation}
The separation of $\v{B}$ in Eq. (\ref{eq:synthetic_magnetic_field_two_parts}), into body-type and stress-type fields, also applies to the vorticity, leading to the following relation for the vorticity tensor: 
\begin{equation}
  \Omega_{ik}=\frac{\alpha}{2m}\epsilon_{ijk}b_j+\frac{\gamma}{2m\rho}\left( A_i\nabla_k\rho-A_k\nabla_i\rho \right).
  \label{eq:vorticity_tensor_2}
\end{equation}
Substituting Eqs. (\ref{eq:vorticity_tensor_2}) and (\ref{eq:scalar_potential_due_to_nonlinear_potentials}) into Eq. (\ref{eq:synthetic_lorentz_force_2}), we notice the double role played by the stress-like vortical field depicted by the two last terms in Eq. (\ref{eq:vorticity_tensor_2}).
The second of these cancels the ``convective'' force contribution from the last term in Eq. (\ref{eq:synthetic_electric_field_2}), while the first combines with $-\nabla_k\phi$ to produce a force which is the divergence of the rank-two tensor 
\begin{equation}
  \Gamma_{jk}=\delta_{jk}\left[ \left( 1-\chi \right)\rho\eta+\gamma J_iA_i \right],
  \label{eq:stress_tensor_nonlinear_potentials}
\end{equation}
such that
\begin{equation}
  \lambda_k=\gamma A_k\delta_{ij}e_{ij}+\alpha\epsilon_{ijk}v_ib_j+\frac{1}{\rho}\nabla_j\Gamma_{jk}.
  \label{eq:synthetic_lorentz_force_3}
\end{equation}
In other words, $\Gamma_{jk}$ characterises the fluid stress brought about by the nonlinear potentials.
However, $\Gamma_{jk}$ is not the full stress tensor of the fluid.
The quantum potential defined in Eq. (\ref{eq:quantum_potential}), gives rise to two additional stress terms, since a little manipulation reveals that $f_k^Q=-\nabla_k Q$ can be derived from a rank-2 tensor $Q_{jk}$:
\begin{equation}
  \rho f^Q_k=\nabla_j Q_{jk},
  \label{eq:force_quantum_potential_tensor_form}
\end{equation}
where, recalling Eq. (\ref{eq:stress_tensor_pressure_viscocity}), we decompose $Q_{jk}$ into the form
\begin{equation}
  Q_{jk}=-P_Q\delta_{jk}+\sigma_{jk}.
  \label{eq:stress_tensor_of_quantum_potential}
\end{equation}
The first contribution on the right hand side of this equation, represents the quantum pressure 
\begin{equation}
  P_Q=-\frac{\hbar^2}{4m}\nabla^2\rho,
  \label{eq:quantum_pressure}
\end{equation}
while the second, is the quantum stress tensor
\begin{equation}
  \sigma_{jk}=-\frac{\hbar^2}{4m\rho}\nabla_j\rho\nabla_k\rho.
  \label{eq:quantum_stress_tensor}
\end{equation}
It is $\sigma_{jk}$ which is responsible for the osmotic pressure driving quantum diffusion \cite{nelson1966derivation,wyatt2006quantum,faris2014diffusion}.
By introducing the osmotic velocity field \cite{wyatt2006quantum}
\begin{equation}
  w_k=-\frac{D}{\rho}\nabla_k\rho,
  \label{eq:osmotic_velocity}
\end{equation}
with diffusion coefficient $D=\hbar/\left( 2m \right)$, Eq. (\ref{eq:quantum_stress_tensor}) takes the form
\begin{equation}
  \sigma_{jk}=-m\rho w_jw_k.
  \label{eq:quantum_stress_tensor_2}
\end{equation}
After substituting Eqs. (\ref{eq:synthetic_lorentz_force_3}) and (\ref{eq:force_quantum_potential_tensor_form}) into Eq. (\ref{eq:Cauchy_equation_2}), a Cauchy equation  
\begin{equation}
  m\rho\frac{dv_k}{dt}=\rho f_k+\nabla_j \Pi_{jk},
  \label{eq:Cauchy_equation_QHJ_2}
\end{equation}
is obtained for the fluid, where the body-forces read
\begin{equation}
  f_k=-\nabla_kV+\gamma A_k e_{ij}\delta_{ij}+\alpha\epsilon_{ijk}v_i b_j,  \label{eq:body_forces_single_component_gauge_potential}
\end{equation}
and the stress tensor of the fluid, is given by 
\begin{equation}
  \Pi_{jk}=-\left[-\frac{\hbar^2}{4m}\nabla^2\rho+\left( \chi-1 \right)\rho\eta-\gamma A_iJ_i\right] \delta_{jk}+\sigma_{jk}.
  \label{eq:stress_tensor_single_component_nonlinear_fluid}
\end{equation}
Since the fluid pressure can be read from the diagonal components of the stress tensor (see Eq. \ref{eq:stress_tensor_pressure_viscocity}), we see that
\begin{equation}
  P=P_Q+\left( \chi-1 \right)\rho\eta-\gamma A_iJ_i,
  \label{eq:pressure_single_component_nonlinear_fluid}
\end{equation}
depends on the overlap of the current density and the vector potential, and as such, depends explicitly on the canonical flow $\v{u}$ of the fluid.
In other words, the fluid pressure becomes a function of both independent dynamical variables $\rho$ and $\v{u}$ and as a consequence, transforms generally from one Galilean frame of reference to another.
Complementing this pressure term, two additional nonlinear body-forces enter Eq. (\ref{eq:body_forces_single_component_gauge_potential}) as a result of $\v{A}\left( \rho \right)$. 
The last term in this expression, represents an $\alpha$-modulated magnetic-like force due to the spatial dependence of $\v{a}$.
The second term on the other hand, results from the time-dependence of $\v{A}$ in Eq. (\ref{eq:continuity_gauge_potential}) and, interestingly, constitutes a body-force of dilation.
This follows from the continuity of fluid mass from Eq. (\ref{eq:continuity}), which can be given the form
\begin{equation}
  e_{ij}\delta_{ij}=-\frac{1}{\rho}\frac{d\rho}{dt}.
  \label{eq:continuity_dilation}
\end{equation}
Therefore, if we track an infinitesimal volume element of fluid as it flows, an additional body-force is exerted throughout the element of the fluid whenever the size of the volume changes.
If for instance, the element is compressed by external fields or as a result of entering an increasing surrounding local pressure field, flow is imparted onto the whole element by the vector potential.
This change in mechanical momentum must be the result of a body-type force. 

\section{\label{sec:multi_component_gauge_potential}Multi-component gauge potential}
So far, we have considered nonlinear vector potentials having a single dynamical component, the orientation of $\v{A}$ at each point of space being fixed in time.
From a mathematical perspective, this restriction can be circumvented by simply allowing for the possibility of additional dynamical components and a multi-component gauge potential to emerge in the form
\begin{equation}
  \v{A}=\sum_{n=1}^N\alpha'_n\left( \rho \right)\v{a}'_n,
  \label{eq:multi_component_gauge_potential}
\end{equation}
where the sum is carried over the total number of static vector fields $\v{a}'_n$ determining the directions of gauge-flow imposed on the system.
Although the $\v{a}'_n$ are fixed and externally prescribed, the orientation of $\v{A}$ will generally change in time through the $\rho$-dependence of the $\alpha'_n$, so long as it is possible to find at least two independent linear combinations from the set $\v{a}'_n$ which have different associated $\rho$-dependent component functions.
When the $\v{a}'_n$ take on values in physical space, the rank $r$ of the matrix of coefficients $a_{ij}\left( \v{x} \right)$ is $r\left( \v{x} \right)\leq 3$.
An orthonormal local basis $a_{ij}\left( \v{x} \right)=\uv{x}_i\cdot\uv{a}_j$ may be constructed for $\v{A}$, such that
\begin{equation}
  A_i=a_{ij}\alpha_j\left( \rho \right),
  \label{eq:multi_component_gauge_potential_matrix}
\end{equation}
where $\alpha_i=\alpha_i\left( \alpha'_1, \cdots, \alpha'_N \right)$, and the $\abs{\v{a}_i}$ have been absorbed into the $\alpha_i$ so as to ensure $\v{a}_i=\uv{a}_i$.
Since $a_{ij}a_{jk}=\delta_{ik}$, the $\alpha_i$ may be inverted as a function of the $A_i$, according to 
\begin{equation}
  \alpha_i=a_{ij}A_j.
  \label{eq:component_functions}
\end{equation}
Notice how each independent vector $\v{A}_i$ comprising the multi-component gauge potential $\v{A}=\uv{x}_iA_i=\uv{a}_i\alpha_i$, gives rise to an associated $\gamma$ from Eq. (\ref{eq:gamma_dimensionless_variable}).
As a consequence, it will be useful to define the functions
\begin{equation}
  \gamma_i=\frac{\rho}{\alpha_i}\frac{d\alpha_i}{d\rho},
  \label{eq:gamma_dimensionless_variable_components}
\end{equation}
not summed over $i$.
In the case where the $\alpha_i$ are identical functions of $\rho$, the multi-component gauge potential is abridged to the single component potential $A_i=a_i\alpha\left( \rho \right)$.
In turn, when the orientation of the local basis is independent of position, $a_{ij}$ can always be reduced to the identity matrix by performing a suitable orthogonal transformation.
This is the multi-component equivalent of a uniform field $\v{a}$ in the single-component system.
In the more general case where $a_{ij}$ depends on position, each basis vector $\v{a}_i$ gives rise to an Eq. (\ref{eq:magnetic_body_field_single_component}), calling for the extension
\begin{equation}
  b_{kp}=\epsilon_{ijk}\nabla_ia_{jp}.
  \label{eq:magnetic_body_field_multi_component}
\end{equation}
\subsection{Dynamics of the condensate}
The dynamics of the field $\theta$ are governed by the canonical field Eq. (\ref{eq:canonical_equations_hydrodynamic_conjugate_theta}).
The Hamiltonian density of the field again takes the form of Eq. (\ref{eq:Hamiltonian_density_nonlinear_quantum_fluid}), but the components of the velocity field, now read
\begin{equation}
  v_i=u_i-\frac{1}{m}a_{ij}\alpha_j\left( \rho \right),
  \label{eq:velocity_field_multi_component_gauge_covariant_polar_form}
\end{equation}
where $u_i=\nabla_i\theta/m$ is the canonical flow. 
In this instance, Eq. (\ref{eq:canonical_equations_hydrodynamic_conjugate_theta}) yields the QHJE 
\begin{equation}
  \dot{\theta}+\frac{1}{2}mv^2-v_i\gamma_{ij}A_j+\chi\eta+V+Q=0,
  \label{eq:QHJ_multi_component_nonlinear_gauge_potential}
\end{equation}
where $\chi$ is again given by Eq. (\ref{eq:chi_dimensionless_variable}), but the dimensionless overlap modulation function $\gamma$ regulating the strength of the nonlinear flow term, generalises to
\begin{equation}
  \gamma_{ij}=\sum_k a_{ik}\gamma_k a_{kj},
  \label{eq:gamma_dimensionless_variable_matrix}
\end{equation}
where the $\gamma_k$ are given by Eq. (\ref{eq:gamma_dimensionless_variable_components}).
The $\gamma_{ij}$ define the couplings which take place in the wave equation between the components of flow $v_i$ and the components of the gauge potential $A_j$ from Eq. (\ref{eq:multi_component_gauge_potential_matrix}). 
Since the basis for $\v{v}$ and the basis for $\v{A}$ are different, the latter depending on position, the matrix of coefficients $\gamma_{ij}$ at a given point of space will not be diagonal in general.
If at point $\v{x}_0$ both bases do coincide, $\gamma_{ij}\left( \v{x}_0 \right)$ is a diagonal matrix with diagonal elements $\gamma_k$.
\subsection{\label{sec:Cauchy_field}Momentum-transport from the field description of the condensate}
In section \ref{sec:momentum_transport_single_component}, a Cauchy equation was derived by observing that the gradient of the QHJE leads to an expression for the convective derivative of the velocity field in Eq. (\ref{eq:Cauchy_equation_single_component_nonlinear_gauge_potential_long}).
However, a more concise route to Cauchy's equation is furnished by the field description of the condensate developed in section \ref{sec:canonical_formalism} (see \cite{holland1995quantum}).
In this treatment, the dynamical state of the matter-field is completely specified by the stress-energy-momentum tensor
\begin{equation}
T_{\mu\nu}=-\sum_{\phi=\rho,\theta}\pd{\mathcal{L}}{(\partial_{\mu}\phi)}\partial_{\nu}\phi +\delta_{\mu\nu}\mathcal{L},
  \label{eq:stress_energy_momentum_tensor}
\end{equation}
while the transport equations governing the dynamics of energy-flow and momentum-flow, follow from the conservation law
\begin{equation}
  \partial_{\mu}T_{\mu\nu}=\partial_{\nu}\mathcal{L},
  \label{eq:conservation_energy_momentum}
\end{equation}
where we have adopted a relativistic-like notation with $\mu=0,1,2,3$. 
Recalling that the passage between the Lagrangian and Hamiltonian descriptions is provided by Eq. (\ref{eq:Lagrangian_density_Hamiltonian_density}), the field Lagrangian density of a Bose-condensed quantum fluid subject to nonlinear potentials, reads 
\begin{equation}
  \mathcal{L}=-\rho\left( \dot{\theta}+\frac{1}{2}mv^2+\eta +V\right) -\mathcal{Q}.
  \label{eq:Lagrangian_density_nonlinear_quantum_fluid}
\end{equation}
Alternatively, $\mathcal{L}$ can be cast in terms of the fields and their spatial derivatives, by inserting the wave equation for $\theta$ into the above expression.
Rendering $\mathcal{L}$ into this form is essential for evaluating the components of the stress tensor $T_{ij}$.
For the particular case (\ref{eq:velocity_field_multi_component_gauge_covariant_polar_form}) considered here, substituting the wave equation (\ref{eq:QHJ_multi_component_nonlinear_gauge_potential}) for $\theta$ into Eq. (\ref{eq:Lagrangian_density_nonlinear_quantum_fluid}), yields
\begin{equation}
  \mathcal{L}=-\frac{\hbar^2}{4m}\nabla^2\rho+\left( \chi-1 \right)\rho\eta-J_i\gamma_{ij}A_j.
  \label{eq:Lagrangian_multi_component_nonlinear_gauge_potential_polar_form_simplified}
\end{equation}

The stress-energy-momentum tensor in Eq. (\ref{eq:stress_energy_momentum_tensor}) characterises the dynamical state of the field by specifying the energy density, the momentum density, and the currents associated with both of these quantities.
The energy density of the field is
\begin{equation}
  -T_{00}\equiv\mathcal{H}.
  \label{eq:energy_density}
\end{equation}
The energy current density $-T_{k0}\equiv \mathcal{S}_k$, takes the form
\begin{equation}
  \mathcal{S}_k=D\dot{\rho}w_k-\rho \dot{\theta}v_k, 
  \label{eq:energy_current_density}
\end{equation}
where $w_k$ is the osmotic velocity from Eq. (\ref{eq:osmotic_velocity}) and $D$ is the quantum diffusion coefficient. 
The canonical momentum density $T_{0k}\equiv \mathcal{P}_k$, reads
\begin{equation}
  \mathcal{P}_k=\rho\nabla_k\theta=\rho m u_k
  \label{eq:canonical_momentum_density}.
\end{equation}
Using both expressions (\ref{eq:Lagrangian_density_nonlinear_quantum_fluid}) and (\ref{eq:Lagrangian_multi_component_nonlinear_gauge_potential_polar_form_simplified}) for $\mathcal{L}$, the canonical momentum current density or stress tensor $T_{jk}$ of the field, is found to be
\begin{equation}
  T_{jk}=\rho m\left( w_jw_k+v_ju_k \right)+\delta_{jk}\mathcal{L}.
  \label{eq:stress_tensor_field}
\end{equation}
The interpretation of Eqs. (\ref{eq:energy_current_density}) and (\ref{eq:stress_tensor_field}) as the respective current densities of the quantities defined in Eqs. (\ref{eq:energy_density}) and (\ref{eq:canonical_momentum_density}), follows from the conservation law in Eq. (\ref{eq:conservation_energy_momentum}), which separates out into an equation of continuity of energy
\begin{equation}
  \partial_t\mathcal{H}+\div{\v{S}}=-\pd{\mathcal{L}}{t},
  \label{eq:transport_energy}
\end{equation}
and an equation of continuity of momentum
\begin{equation}
  \partial_t \mathcal{P}_k+\nabla_j T_{jk}=\pd{\mathcal{L}}{x_k},
  \label{eq:transport_momentum}
\end{equation}
which is simply the field equivalent of Cauchy's equation (\ref{eq:Cauchy_equation}).
Note that the right hand sides of Eqs. (\ref{eq:transport_energy}) and (\ref{eq:transport_momentum}) should be evaluated holding the fields and their derivatives constant.
Notice also the difference in sign convention used for the fluid stress $\Pi_{jk}$ in Eq. (\ref{eq:Cauchy_equation}) and the field stress $T_{jk}$ in Eq. (\ref{eq:transport_momentum}).
In addition to this sign difference, $\Pi_{jk}$ and $T_{jk}$ differ by a flow-stress term $m\rho v_ju_k$ as a result of the relative motion between the fluid and field reference frames, such that
\begin{equation}
  T_{jk}=-\Pi_{jk}+m\rho v_ju_k.
  \label{eq:stress_tensor_relation_moving_frame}
\end{equation}
Substituting Eqs. (\ref{eq:canonical_momentum_density}) and (\ref{eq:stress_tensor_field}) into (\ref{eq:transport_momentum}) and making use of the continuity of fluid mass, leads to the following canonical momentum-transport equation for the fluid:
\begin{equation}
  m\rho\left( \pd{}{t}+\v{v}\cdot\grad{} \right)u_k=\pd{\mathcal{L}}{x_k}+\nabla_j\Pi_{jk},
  \label{eq:Cauchy_equation_canonical_momentum_1}
\end{equation}
where the fluid stress takes the form
\begin{equation}
  \Pi_{jk}=-P\delta_{jk}+\sigma_{jk},
  \label{eq:stress_tensor_field_2}
\end{equation}
with $\sigma_{jk}$ denoting the quantum stress tensor from Eq. (\ref{eq:quantum_stress_tensor}), and $P$ the fluid pressure, given by 
\begin{equation}
  P=P_Q+\left( \chi-1 \right)\rho\eta-J_i\gamma_{ij}A_j.
  \label{eq:pressure_multi_component_nonlinear_fluid}
\end{equation}
This highlights the equivalence of the fluid pressure and the field Lagrangian density from Eq. (\ref{eq:Lagrangian_multi_component_nonlinear_gauge_potential_polar_form_simplified}). 
Supplementing the fluid stress, Eq. (\ref{eq:transport_momentum}) indicates that a current 
\begin{equation}
  \pd{\mathcal{L}}{x_k}=\rho \left(-\nabla_k V+\alpha_j v_i\nabla_k a_{ij}\right)
  \label{eq:current_injection_multi_component_gauge_potential}
\end{equation}
is injected into the field as a result of the generally position-dependent local basis $a_{ij}$.
The transport of canonical momentum in the fluid frame of reference is completely determined by the fluid stress and the body force density from Eqs. (\ref{eq:stress_tensor_field_2}) and (\ref{eq:current_injection_multi_component_gauge_potential}) respectively.
A mechanical momentum-transport equation may then be obtained, simply by substituting the canonical flow $u_k$ in Eq. (\ref{eq:Cauchy_equation_canonical_momentum_1}) for the mechanical flow $v_k$ from Eq. (\ref{eq:velocity_field_multi_component_gauge_covariant_polar_form}). 
In other words, the difference between canonical and mechanical momentum-transport stems from the additional body forces generated by the time dependence and the spatial dependence along fluid streamlines of the gauge potential, these taking the form
\begin{equation}
  -\frac{dA_k}{dt}=\gamma_{kn}A_n\delta_{ij}e_{ij}-\alpha_jv_i\nabla_ia_{kj}.
  \label{eq:time_dependence_multi_component_gauge_potential}
\end{equation}
Equations (\ref{eq:velocity_field_multi_component_gauge_covariant_polar_form}), (\ref{eq:current_injection_multi_component_gauge_potential}) and (\ref{eq:time_dependence_multi_component_gauge_potential}), combine in expression (\ref{eq:Cauchy_equation_canonical_momentum_1}) to yield a Cauchy equation (\ref{eq:Cauchy_equation}) for the fluid, where the stress tensor of the fluid is given by Eq. (\ref{eq:stress_tensor_field_2}) and the body force of the single component case, generalises to 
\begin{equation}
  f_k=-\nabla_kV+\gamma_{kn}A_n\delta_{ij}e_{ij}+\alpha_n\epsilon_{ijk}v_ib_{jn},
  \label{eq:body_forces_multi_component_gauge_potential}
\end{equation}
where $b_{jn}$ is given by Eq. (\ref{eq:magnetic_body_field_multi_component}).
\\
\section{\label{sec:conclusion}CONCLUSION}

The hydrodynamic canonical formalism is an ideal framework for investigating the dynamics of a condensate matter-field whose effective Hamiltonian features nonlinear interaction terms which take the simplest form when expressed as functionals of the amplitude and the phase of the complex field.
For instance, when the effective kinetic energy of the field becomes a nonlinear functional of the density by means of a density-dependent gauge potential, it becomes easy to see that nonlinear flow-dependent terms invariably enter the wave equation of the condensate. 
In turn, two non-trivial terms emerge in the mechanical momentum-transport equation of the fluid: a flow-dependent fluid pressure and a body force of dilation.
These should have important consequences for both the Galilean and gauge symmetries of the fluid, where new transformations laws may be required in order to restore the invariance of the fluid under the transformation groups.
The immediate lack of Galilean invariance should also carry significant implications for the elementary excitations of the fluid.
For instance, it should no longer be the case that the velocity of sound $\v{v}_s$ be determined simply by $\partial P/\partial\rho$, since $P$ depends explicitly on the flow.
This calls for a generalisation of the expression used to derive the $\v{v}_s$ of a fluid in terms of the particular dependence of the fluid pressure on the density.
Finally, the nonlinear body force of dilation will appear in the expectation value of the time derivative of the mechanical momentum and could therefore be investigated numerically in the drag force acting on an impurity moving through the fluid. 
For typical quantum fluids, the drag force is determined by the configuration of the fluid density in the vicinity of the localised object potential.
In contrast, the reaction to the body force of dilation exerted on a fluid subject to $\v{A}\left( \rho \right)$, should occur throughout the whole fluid, taking place wherever current and density gradients overlap.
We expect the onset of vortex nucleation to depend on the relative orientation of the gauge potential and the impurity velocity, and the flow-dependent pressure to play an important role in pressure drag.

\subsection*{Acknowledgements}

We would like to thank Lawrence Phillips and Manuel Valiente for helpful and interesting discussions.
Y.B acknowledges support from EPSRC CM-CDT Grant No. EP/L015110/1.
P.{\"O} acknowledges support from EPSRC grant No. EP/M024636/1.

%

\end{document}